\documentclass[twocolumn,showkeys,aps,prb,showpacs]{revtex4-1}
\usepackage{graphicx}
\usepackage[CJKbookmarks,dvipdfm,colorlinks,linkcolor=blue,citecolor=blue]{hyperref}

\begin{document}

\title{Strain effects on phonon transport in antimonene from a first-principles study}

\author{Ai-Xia Zhang, Jiang-Tao Liu and San-Dong Guo}
\affiliation{School of Physics, China University of Mining and
Technology, Xuzhou 221116, Jiangsu, China}

\begin{abstract}
Strain engineering is a very  effective method  to continuously tune the electronic,  topological, optical and thermoelectric properties of materials. In this work, strain-dependent phonon transport of recently-fabricated antimonene (Sb monolayer) under  biaxial strain is investigated  from a combination of  first-principles calculations and the linearized phonon Boltzmann equation within the single-mode relaxation time approximation (RTA).  It is found that the ZA dispersion of antimonene
with strain less than -1\%  gives imaginary frequencies, which suggests that compressive strain can induce structural instability. Experimentally, it is possible to enhance structural stability by tensile strain. Calculated results show that lattice thermal conductivity increases with strain changing from -1\% to 6\%, and lattice thermal conductivity at 6\% strain
is 5.6 times larger  than that at -1\% strain at room temperature. It is interesting that lattice thermal conductivity  is in inverse proportion to buckling parameter $h$ in considered strain range. Such a  strain dependence  of  lattice thermal conductivity is attributed to enhanced  phonon lifetimes caused by increased strain, while group  velocities have a decreased effect on lattice thermal conductivity with increasing strain. It is found that acoustic branches dominate the lattice thermal conductivity over the full strain range. The cumulative room-temperature lattice thermal conductivity at -1\% strain
converges  to  maximum  with phonon mean free path (MFP) at 50 nm, while one at 6\% strain
becomes as large as 44 $\mathrm{\mu m}$, which suggests that strain can give rise to very  strong size effects on
lattice thermal conductivity in antimonene. These results may provide guidance  on fabrication
techniques of antimonene, and offer perspectives on tuning lattice thermal conductivity by size and strain for applications of
thermal management and thermoelectricity.
\end{abstract}

\pacs{72.15.Jf, 71.20.-b, 71.70.Ej, 79.10.-n ~~~~~~~~~~~~~~~~~~~~~~~~~~~~~~~~~~~Email:guosd@cumt.edu.cn}

\maketitle

\section{Introduction}
Due to  their novel physical and chemical properties,  two-dimensional (2D) materials have been widely investigated both theoretically and experimentally, including semiconducting transition-metal dichalcogenide\cite{q7}, group IV-VI\cite{q8}, group-VA\cite{q9,q10}  and group-IV\cite{q11}  monolayers. Graphene, as representative of these monolayers,  has a unique massless Dirac-like electronic excitation\cite{t1}, and has extremely
high lattice  thermal conductivity\cite{t2}.
In comparison with the gapless graphene, $\mathrm{MoS_2}$ is another classic 2D monolayer  with intrinsic direct band gap of 1.9 eV, which  has been applied in field effect transistors, photovoltaics and photocatalysis\cite{t3,t4,t5,t6}.
Recently, semiconducting group-VA elements (As, Sb, Bi) monolayers with the  graphene-like buckled structure are predicted to be of good stability  by the first-principle calculations\cite{q9}. Subsequently,  Sb monolayer (antimonene) of them  is  successfully  synthesized on various substrates via van der Waals epitaxy growth\cite{t8,t9}.  Experimentally, graphene-like buckled structure is confirmed for antimonene  by Raman spectroscopy and transmission electron microscopy\cite{t8,t9}.
\begin{figure}
  \includegraphics[width=5.5cm]{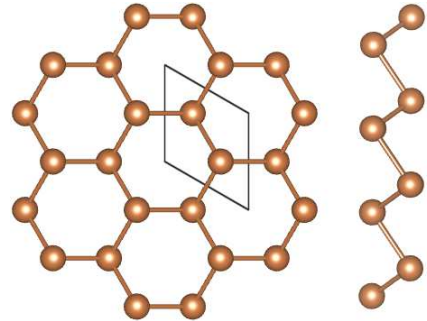}
  \caption{(Color online) Top and side view of the crystal structure
of antimonene,  and the frame surrounded by a black box is  unit cell.}\label{st}
\end{figure}
\begin{figure}
  \includegraphics[width=8cm]{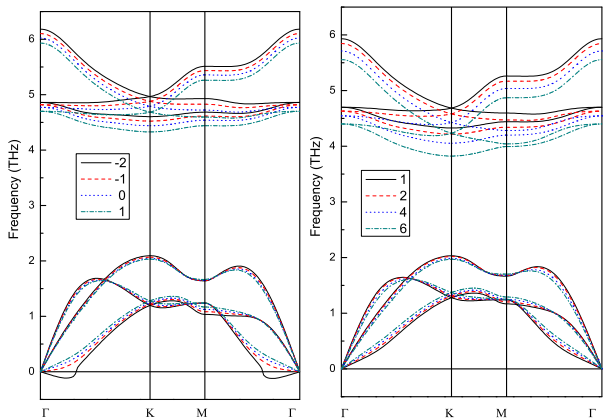}
  \caption{(Color online) Phonon band structures of antimonene with strain from  -2\% to 6\%.}\label{ph}
\end{figure}

\begin{figure}[!htb]
  \includegraphics[width=8cm]{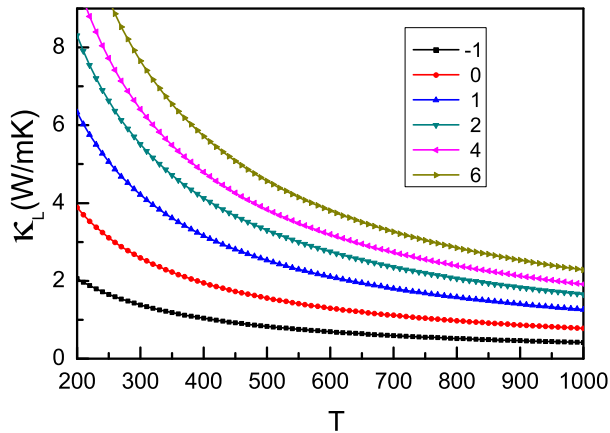}
  \caption{(Color online) The lattice thermal conductivities  of antimonene with strain from  -1\% to 6\% as a function of temperature.}\label{kl}
\end{figure}
In theory, thermal transport of  antimonene has been widely investigated, including electron and phonon parts\cite{l1,l2,l3,l4}.
It has been predicted that chemical functionalization can  effectively reduce lattice thermal conductivity of antimonene\cite{l3}. The lattice thermal conductivity of antimonene  is predicted to be  15.1 $\mathrm{W m^{-1} K^{-1}}$\cite{l1}, 13.8 $\mathrm{W m^{-1} K^{-1}}$\cite{l2} and 2.3 $\mathrm{W m^{-1} K^{-1}}$\cite{l3} based on first-principles calculations and the phonon Boltzmann equation.  A remarkable discrepancy may be because the different thickness of antimonene is adopted to calculate lattice thermal conductivity.
Recently, we  have systematically investigated the thermoelectric properties of group-VA elements (As, Sb, Bi) monolayers\cite{l4}, and the  lattice thermal conductivity from As to Bi monolayer decreases  by using sheet thermal conductance for a fair comparison. Strain can effectively tune the intrinsic physical properties of 2D materials, such as  electronic and thermoelectric properties. Semiconductor-metal phase transition can be induced in $\mathrm{MoS_2}$ monolayer by applied strain, and strain can also lead to  significantly enhanced power factor\cite{l5,l6}. For $\mathrm{PtSe_2}$ and $\mathrm{ZrS_2}$,
tensile strain can induce reduced lattice thermal conductivity\cite{l7,l8}.  Tensile strains can induce strong size effects for
lattice thermal conductivities of silicene, germanene and stanene, and their lattice thermal conductivities  firstly increase with increasing strain, and then decrease  with further strain\cite{l9,l10}.  Phonon  transport in both strained and unstrained graphene has been investigated by a rigorous first principles Boltzmann-Peierls equation, and the out-of-plane acoustic branch  provides the dominant contribution to lattice thermal conductivity of graphene\cite{l100}. 
In real applications,  the  residual strain usually exists in  nanoscale devices\cite{l11}, and the conduction of heat plays a important role  in real-world devices. Therefore, it is interesting and necessary to investigate strain effects on phonon transport in recently-fabricated antimonene.

\begin{figure}[!htb]
  \includegraphics[width=8cm]{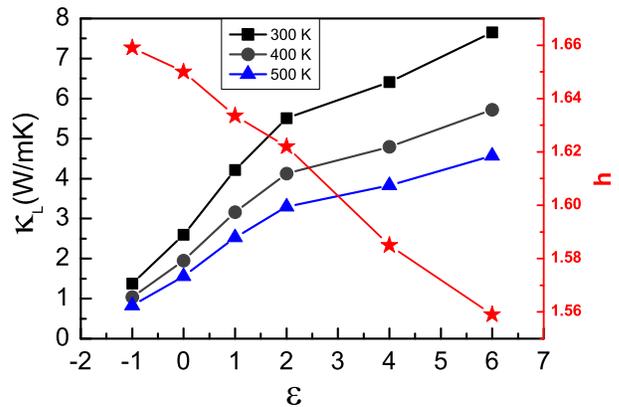}
  \caption{(Color online) The lattice thermal conductivities (300, 400 and 500 K)  and  buckling parameter $h$  ($\mathrm{{\AA}}$) of antimonene  versus  strain $\varepsilon$.}\label{kl1}
\end{figure}

In this work, the strain-dependent phonon transport in antimonene is investigated based on the single-mode
RTA. It is found that  compressive strain can produce structural instability in  antimonene, while tensile strain can  enhance structural stability. The lattice thermal conductivity increases with increasing strain, which  is in inverse proportion to buckling parameter $h$.  The tensile-strain enhanced lattice thermal conductivity is due to improved phonon lifetimes caused by increased strain.  The strain can induce  very  strong size effects on
lattice thermal conductivity in antimonene, indicating the possibility  to
manipulate  thermal management by strain and size.

The rest of the paper is organized as follows. In the next
section, we shall give our computational details about  the first-principle and  phonon transport calculations. In the third section, we shall present strain effect on  phonon transport of antimonene. Finally, we shall give our  conclusions in the fourth section.

\begin{figure*}
  \includegraphics[width=15cm]{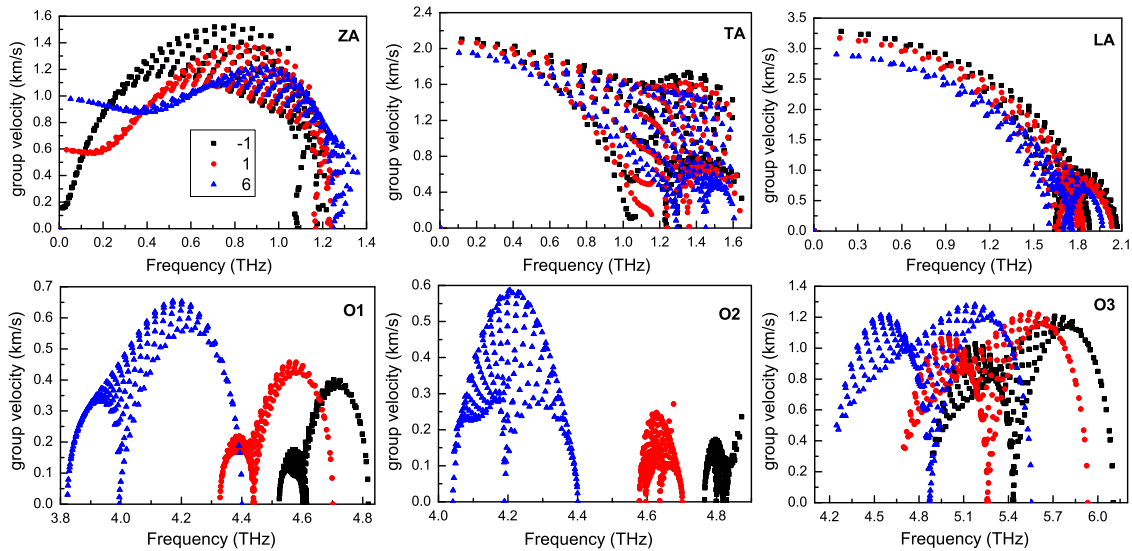}
  \caption{(Color online) The phonon mode group velocities of antimonene with strain being  -1\%, 1\% and 6\% in the first Brillouin zone.}\label{v}
\end{figure*}

\section{Computational detail}
First-principles calculations are performed within projector augmented-wave method  and
 the generalized gradient approximation (GGA) exchange-correlation functional of Perdew-Burke-Ernzerhof (PBE),
as implemented in the VASP code\cite{pv1,pv2,pv3,pbe}. The unit cell  of antimonene   is built with the vacuum region of larger than 16 $\mathrm{{\AA}}$ to avoid spurious interaction. The
electronic stopping criterion is $10^{-8}$ eV.
The  lattice thermal conductivity of antimonene   is carried out with the single mode RTA and linearized phonon Boltzmann equation  using Phono3py code\cite{pv4}.
The second order harmonic and third
order anharmonic interatomic force constants (IFC)  are
calculated by using a 5 $\times$ 5 $\times$ 1  supercell  and a  4 $\times$ 4 $\times$ 1 supercell, respectively.
Using the harmonic IFCs, phonon dispersion can be attained by Phonopy package\cite{pv5}, which determines the allowed three-phonon scattering processes, and further the  group velocity  and specific heat can be calculated.
 Based on third-order anharmonic IFCs, the three-phonon scattering rate can be calculated, which determines  the phonon lifetimes. To compute lattice thermal conductivities, the
reciprocal spaces of the primitive cells  are sampled using the 50 $\times$ 50 $\times$ 2 meshes.
For 2D material, the calculated  lattice  thermal conductivity  depends on the length of unit cell used in the calculations along z direction\cite{2dl}.  The lattice  thermal conductivity should be normalized by multiplying Lz/d, in which  Lz is the length of unit cell along z direction  and d is the thickness of 2D material, but the d  is not well defined.   In this work, the length of unit cell (18 $\mathrm{{\AA}}$) along z direction is used as the thickness of  antimonene. To make a fair comparison between various 2D monolayers, the thermal sheet conductance can be used, defined as $\kappa$ $\times$ d.

\section{MAIN CALCULATED RESULTS AND ANALYSIS}
The antimonene ($\beta$-phase) possesses a graphenelike buckled honeycomb structure\cite{q9,t8,t9}, which is shown in \autoref{st}.
Similar  structure configuration can be found in silicene, germanene, and stanene with a small buckling\cite{l9,l10}.
The $\varepsilon=(a-a_0)/a_0$ is defined to describe biaxial strain, in which $a_0$ is the unstrain lattice constant.  The calculations are  performed  on  nine values of $\varepsilon$ ranging from -6\% to 6\%, and the $\varepsilon$$<$0 ($>$0) means compressive (tensile) strain. Firstly, we  calculate the phonon dispersion in high symmetry directions with strain from -6\% to 6\%. Calculated results show that the phonon dispersions give imaginary frequencies with strain less than -1\%, which indicates
that  these freestanding antimonene are not stable. These results may suggest that tensile strain can strengthen the stability of antimonene in experiment.  Similar results also can be found in germanene, and stanene, and tensile strain can harden soft modes\cite{l10}. Here, only phonon dispersion curves with strain from -2\% to 6\%  are plotted in \autoref{ph}.
Although out-of-plane acoustic modes of antimonene have coupling with the in-plane longitudinal acoustic (LA) and transversal acoustic (TA) modes due to buckling, they are still marked with  ZA  modes.
It is clearly seen that the ZA dispersion at -2\% strain  shows  imaginary frequencies  near the $\Gamma$ point, and the imaginary frequencies become large with increasing compressive strain.
The optical branches overall move downward with increasing  strain from -1\% to 6\%, which can be explained by the reduction of the material stiffness.  These phonon softening for optical branches  are similar to those in
flat graphene and buckled silicene, germanene and stanene\cite{l10}.
The phonon band gap decreases from 2.46 THz to 1.86 THz with strain changing from -1\% to 6\%.
It is found that the slope of the ZA branch increases with increasing strain,  while the ones of LA and TA branches  decrease.
Moreover, the slope of ZA branch  changes more  significantly than ones of LA and TA branches.
It is found that dispersion of LA and TA modes of antimonene changes  slightly
with increasing strain, which is comparable with silicene, germanene, and stanene, but is different from graphene\cite{l9,l10}.

\begin{figure*}
  \includegraphics[width=15cm]{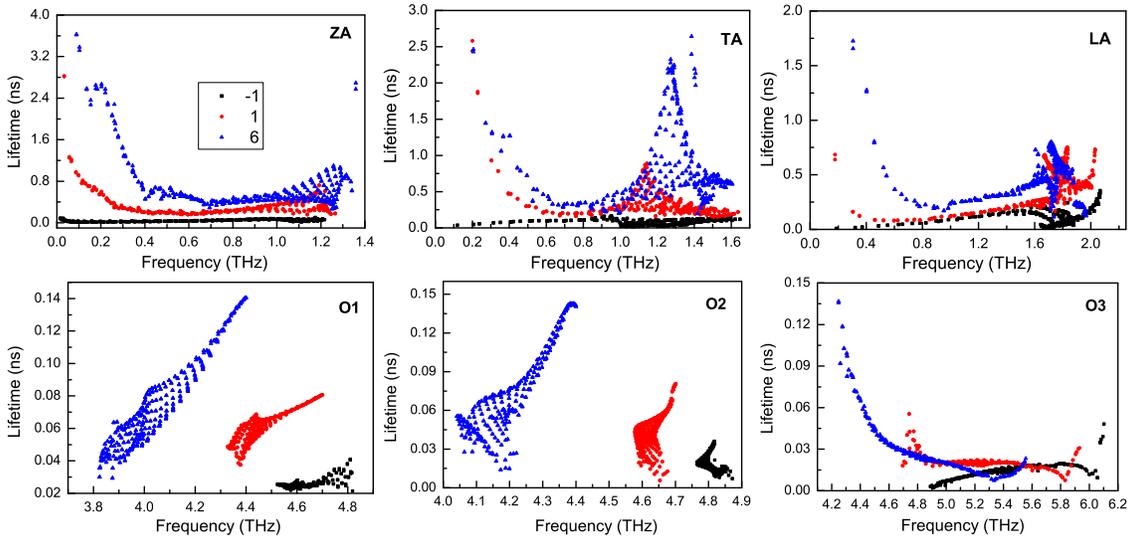}
  \caption{(Color online) The phonon mode lifetimes of antimonene with strain being  -1\%, 1\% and 6\% in the first Brillouin zone.}\label{t}
\end{figure*}
\begin{figure}[!htb]
  \includegraphics[width=8cm]{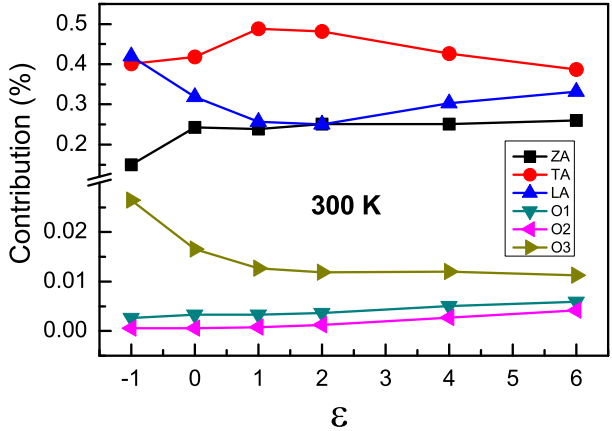}
  \caption{(Color online) phonon modes contributions to total lattice thermal conductivity as a function of strain $\varepsilon$ (300 K).}\label{mode}
\end{figure}

The lattice  thermal conductivities of  antimonene in the strain range of -1\% - 6\% are calculated with RTA method, and they as a function of temperature are plotted \autoref{kl}. Calculated results predict  that the lattice  thermal conductivity increases with strain increasing in considered strain range.  To study changing speed of lattice  thermal conductivity,  the lattice thermal conductivities as a function of strain  at the temperature of  300, 400 and 500 K are shown in \autoref{kl1}.
The lattice thermal conductivity first increases  rapidly with strain being less than 2\%,  and then rises more slowly.
It is predicted that lattice thermal conductivities of  silicene, germanene, and stanene show similar increase with increasing strain within certain strain range\cite{l9,l10}, and the lattice thermal conductivities decrease, when the strain is greater than critical value.
The strain dependence of lattice  thermal conductivity of  antimonene is different from those  of  monolayer $\mathrm{PtSe_2}$ and $\mathrm{ZrS_2}$\cite{l7,l8}.
From As to Bi monolayer,  the lattice thermal conductivity decreases, and the corresponding buckling parameter $h$ increases\cite{l4}.
The lattice  thermal conductivity should have  a  correlation with the buckling parameter, and larger buckling parameter $h$ means lower lattice  thermal conductivity. Therefore, the buckling parameter $h$ as a function of strain is also plotted in \autoref{kl1}. With increasing strain from -1\% to 6\%, the buckling parameter $h$ decreases, and then incremental lattice  thermal conductivity is produced.  Similar relation between  lattice  thermal conductivity and buckling parameter can also be found in  silicene\cite{l9}. This may apply to other
2D monolayers with intrinsic buckling, such as germanene,  stanene, As and Bi monolayers.
Moreover, As-As bond
length keeps increasing with  strain becoming larger. The decreasing buckling parameter $h$ and increasing As-As bond
length mean that  tensile strain can make the structure of antimonene become
more planar.

To identify the underlying mechanism of strain-dependent lattice thermal conductivity, phonon mode group velocities  of antimonene are calculated, using the harmonic IFCs, which  are plotted in \autoref{v}, with -1\%, 1\% and 6\% strains. For ZA branch,
the largest  group velocity  near $\Gamma$ point increases from   0.16  $\mathrm{km s^{-1}}$ to 0.59  $\mathrm{km s^{-1}}$ to  0.98  $\mathrm{km s^{-1}}$ with strain changing from -1\% to 1\% to 6\%.  This can be understood by increased stiffness in Z direction caused by increased strain in X and Y directions. The largest  group velocity of TA (LA) near $\Gamma$ point  decreases
from 2.11 (3.28)  $\mathrm{km s^{-1}}$  for -1\% strain  to 2.07 (3.17)  $\mathrm{km s^{-1}}$  for 1\% strain to 1.95 (2.90) $\mathrm{km s^{-1}}$ for 6\% strain, which is due to weakened As-As interatomic interaction in X and Y directions caused by strain-increased As-As bond length. These results agree well with change of the slope of ZA, TA and LA branches caused by increased strain. Most of group velocities for ZA branch become small with increasing strain, and  group velocities for TA branch
have small changes, and group velocities for LA branch change  slightly small. These strain-dependent group velocities should
lead to decrescent lattice thermal conductivity with increasing strain. The group velocities of optical branches increase with strain increasing, especially for first two optical branches,  which can produce larger lattice thermal conductivity,  but the optical branches have little contribution to total lattice thermal conductivity (hereinafter discussed). Therefore, the changes of  group velocities caused by increasing strain should make  opposite contribution to increased   lattice thermal conductivity.

Next, we consider  strain-dependent  phonon  lifetimes,  which are merely proportional to lattice thermal conductivity in the single-mode RTA method\cite{pv4}.
The phonon lifetimes can be calculated by  three-phonon scattering rate from  third-order anharmonic IFCs.
The  phonon  lifetimes with -1\%, 1\% and 6\% strains are shown in \autoref{t}.
For all phonon branches, most of phonon  lifetimes  become large  with increasing strain, which can induce large lattice thermal conductivity. It is clearly seen that acoustic branches have very larger group velocities and phonon  lifetimes than optical ones, which means that acoustic branches dominate the lattice thermal conductivity. Therefore, we only consider group velocities and phonon  lifetimes of acoustic branches to find out  the underlying mechanism of enhanced  lattice thermal conductivity caused by increased strain. The strain dependence of  group velocities has decreased effect on lattice thermal conductivity, while strain dependence of phonon  lifetimes has enhanced influence. So, the improved phonon lifetimes with increasing strain lead to strain-dependent lattice thermal conductivity. Similar mechanism can be found in silicene, germanene, and stanene, and the acoustic phonon lifetimes determine their strain dependence of lattice thermal conductivity\cite{l9,l10}.
\begin{figure}
  \includegraphics[width=8.0cm]{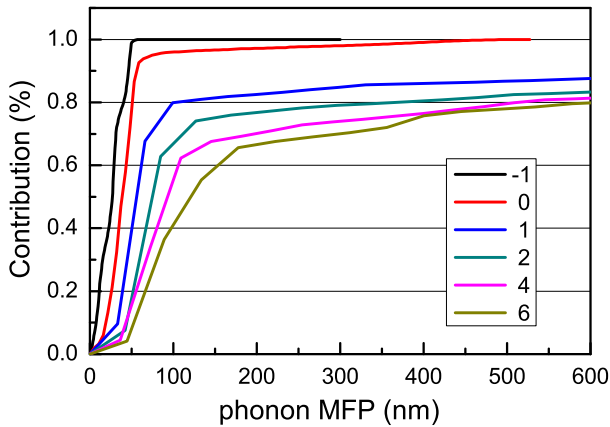}
  \caption{(Color online) Cumulative lattice thermal conductivity divided by total lattice thermal conductivity with respect to phonon mean free path at room temperature.}\label{mkl}
\end{figure}

To further understand strain dependence of lattice thermal conductivity, the  relative contributions of  acoustic and optical branches to the total lattice thermal conductivity as a function
of strain are plotted \autoref{mode}. It is clearly seen that the acoustic modes  give the
dominant contribution, while contribution from optical branches is 1.7\% - 3.0\%.
 In considered strain range, the ZA branch has the smallest contribution among acoustic branches except 2\% strain,  and the TA branch has the largest contribution but -1\% strain. The LA
and TA branches contribute more than 71\% of the total thermal
conductivity,  up to 82\% especially for -1\% strain. This is the same with  silicene\cite{l9}, but
is quite different from graphene, for which  ZA branch  has a
major contribution\cite{l100}.
 The contribution of ZA mode first increases from -1\%  to 0\% strain, and then remains  about the same. The lattice thermal conductivity contributed by the TA mode first increases from -1\%   to 1\% strain, and then decreases. However, the contribution of LA mode   with strain up to 2\%  first
decreases, and then increases.  The  contribution from the third optical branch  first rapidly decreases 
from -1\%  to 2\% strain, and then  slowly decreases  from 2\%  to 6\% strain.  The contributions of  first two optical modes keep  increasing tendency from -1\% to 6\% strain.

To study size effect on lattice thermal conductivity, cumulative lattice thermal conductivity divided by total lattice thermal conductivity with respect to MFP, at room temperature, are  plotted in \autoref{mkl}, with strain from -1\% to 6\%. With MFP increasing, the cumulative lattice thermal conductivity increases, and then approaches maximum.  The corresponding MFP of maximum  changes from 50 nm  to 43590 nm  with strain from -1\% to 6\%. This means that lattice thermal conductivity  is contributed by phonons with longer  MFP, as the strain becomes larger.  The  critical MFP at 6\% strain  is 868 times larger than that at -1\% strain, which demonstrates that strain induces very  strong size effects for lattice
thermal conductivity  of antimonene. The strong size effects on lattice thermal conductivity caused by strain can also be found in  in silicene, germanene, and stanene\cite{l10}. The scale reduction may be an effective method  to reduce lattice thermal conductivity for
antimonene with large strain. When the lattice thermal conductivity is  reduced to 60\% by nanostructures,  the characteristic length changes  from 29 nm to 155 nm  with strain from -1\% to 6\%.

\section{Conclusion}
In summary,   the lattice thermal conductivity of antimonene under
strain is  performed  by the first-principles calculations and  linearized phonon Boltzmann equation within the single-mode RTA.
It is found that increasing strain  can harden and stabilize long wavelength ZA  acoustic branch.  This may provide  guidance on
fabrication of As and Bi monolayers by tensile strain.
From -1\% to 6\% strain, the lattice thermal conductivity of antimonene increases, which is mainly due
to the strain-dependent phonon lifetimes.  The phonon lifetimes increase with increasing
strain, which leads to enhanced lattice thermal conductivity. Strain can induce  a strong size effect on lattice thermal conductivity in antimonene,  which means that the lattice thermal conductivity can be effectively tuned by size and strain. This work provides insight into strain dependence of phonon transport in antimonene, and  As and Bi monolayers  may have similar strain
dependence of lattice thermal conductivity.

\begin{acknowledgments}
This work is supported by the National Natural Science Foundation of China (Grant No. 11404391). We are grateful to the Advanced Analysis and Computation Center of CUMT for the award of CPU hours to accomplish this work.
\end{acknowledgments}

\end{document}